# Simulation of background reduction and Compton depression in low-background HPGe spectrometer at a surface laboratory

NIU Shun-Li(牛顺利)[1;1] CAI Xiao(蔡啸)[1;2] WU Zhen-Zhong(吴振忠)[1] XIE Yu-Guang(谢宇广)[1]
YU Bo-Xiang(俞伯祥)[1] WANG Zhi-Gang(王志刚)[1] FANG Jian(方建)[1] SUN Xi-Lei(孙希磊)[1]
SUN Li-Jun(孙丽君)[1] LIU Ying-Biao(刘颖彪)[12] GAO Long (高龙)[12] ZHANG Xuan (张煊)[12]
ZHAO Hang(赵航)[12] ZHOU Li(周莉)[1] LV Jun-Guang(吕军光)[1] HU Tao(胡涛)[1]
（**1** State Key Laboratory of Particle Detection and Electronics, (Institute of High Energy Physics,Chinese Academy of Sciences) Beijing 100049, China；**2** University of Chinese Academy of Sciences, Beijing 100049, China）

**Abstract**：High-purity germanium detectors are well suited to analysis the radioactivity of samples. In order to reduce the environmental background, low-activity lead and oxygen free copper are installed outside of the probe to shield gammas, outmost is a plastic scintillator to veto the cosmic rays, and an anti-Compton detector can improve the Peak-to-Compton ratio. Using the GEANT4 tools and taking into account a detailed description of the detector, we optimize the sizes of the detectors to reach the design indexes. A group of experimental data from a HPGe spectrometer in using were used to compare with the simulation. As to new HPGe Detector simulation, considering the different thickness of BGO crystals and anti-coincidence efficiency, the simulation results show that the optimal thickness is 5.5cm, and the Peak-to-Compton ratio of $^{40}$K is raised to 1000 when the anti-coincidence efficiency is 0.85. As the background simulation, 15 cm oxygen-free copper plus 10 cm lead can reduce the environmental gamma rays to 0.0024 cps/100 cm$^3$ Ge (50keV~2.8MeV), which is about 10$^{-5}$ of environmental background.
**Key words**：HPGe, Geant4 Simulation, Gamma background, Anti-Compton ratio
**PACS**：29.30.Kv, 29.40.Wk

## 1. Introduction

High-purity germanium (HPGe) detectors are widely used for different experimental researchs, such as neutrino experiments and dark matter experiments. Due to its high energy resolution and efficiency, HPGe detectors are also used to analyze the radioactive of material. IHEP had built a HPGe detector three years ago, used for the low-radioactive materials selected for Daya Bay experiment. But for the future Jiangmen Underground Neutrino Observatory (JUNO) experiments, a more restrict low background experiment, it is beyond the current HPGe's ability, so a ultra low-background HPGe spectrometer is required. It plays a very important role in these experiments to minimize the background, i.e. the natural radioactivity from the lab materials and cosmic ray. To reduce the adverse effects of cosmic rays[1,2], and improve the detection ability of uranium and thorium, there are some general ways, such as moving to underground or cave, using high purity and high density setup materials to shield the gamma, for example, low-background lead and oxygen-free copper. But it is costly and inconvenient. Most of the surface labs, by adding cosmic veto detectors and shielding setup materials to reduce the integral background count-rate of HPGe Spectrometer. It can remove the Compton plateau further reduce the background by placing an anti-Compton detector surround the HPGe probe. It is impossible to completely reject these background, especially the cosmic ray muons for surface labs. The environmental gamma background, with energy below 3 MeV, can be depressed mostly by proper shielding materials, so it is crucial to estimate the accurately dimensions. Based on Geant4 Monte-Carlo simulation, the relatively accurate geometry parameters can be obtained. In this article, we present a new method to simulate the background, a group of experimental data were used to check this method. Further more, the proper dimensions of new HPGe detector such as shielding lead and BGO crystal were optimized.

## 2. A new method to simulate background

Environment has a lot of backgrounds, roughly divided into two parts:
- Cosmic rays. Consisting of muons, neutrons and other nucleonic components.
- α, β and γ rays. Some come from the decay of radioactive elements $^{232}$Th, $^{238}$U, $^{60}$Co, $^{40}$K and so on in the room materials, other are secondary particles from the background and materials interaction. For the α and β rays have short radiation length, gamma rays become the main background.

In this simulation, small part components of cosmic rays, such as neutrons and other nucleonic components were ignored, the material radioactive from the spectrometry was discarded too. Only gamma background of environment and mouns of cosmic rays are considered.

### 2.1 Gamma background simulation

For the surface HPGe detector background simulation, traditional γ background simulation methods in Geant4[3] usually have three steps, firstly, construct the physicslist file, and add the appropriate physical processes in accordance with the purpose of simulation.

The second is to build the geometry, not only the spectrometer but also the layout of the room, the more detailed the geometry, the more accurate the simulation results; The last is the particle generator, which is the most important, usually just natural radioactive of Radon,$^{40}$K, uranium and thorium series of lines are considered. For these simulations, the first two steps are relatively simple, no matter how complex the room is. The third step is the most difficult, it is manifesting in two aspects, the initial position of the radioactive decays and the radioactivity levels of various objects in the room, these data are difficult to be obtained.

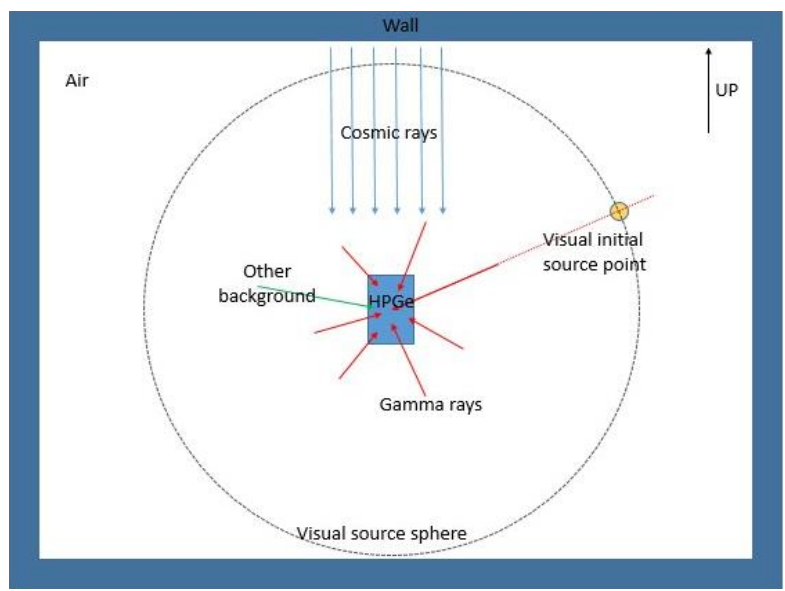

Fig.1 Visual source sphere surround HPGe

Because of the above shortages, a new method is proposed to simulate the gamma-ray background. Imagining a visual source sphere surround the HPGe detector as shown in Fig.1, every gamma entering into the probe can be found a crossing point on the sphere, which is assumed as as the initial decay position. Using the General Particle Source (GPS), another particle gun of Geant4, we can simulate the gamma background without considering the room layout, only initial gamma energy spectrum is needed, which can be obtained by analyzing the background data measured by the HPGe detector. Combining the detection efficiency and energy resolution of the probe, the energy spectrum of virtual particles of spherical shell can be calculated.

## 2.2  Cosmic ray and Radon inside detector simulation

Cosmic rays induce the background arising from the interactions of nucleons and muons with materials surrounding the germanium detector. The muons penetrating the lead and copper shield produce a background in the detector. The muon spectra at sea level were measured in many experiments, but the results are not the same in different regions due to the different geographical environment. Cosmic data measured in Beijing[4,5] was used to generate direction and energy distribution of cosmic ray particles in our case. Radon is a naturally gas with a half-life of 3.8 days that is produced continuously from the natural decay of $^{238}$U, which is spread all over the earth. As a consequence of this, $^{222}$Rn and $^{220}$Rn is the main source of internal radiation exposure to human life, and their decay daughters are also generate background signal. The concentration levels of radon had been been performed in many parts of the world[6,7,8], the results are strongly affected by geological and geophysical conditions. The indoor radon levels of Beijing is about 20 ~ 60 Bq/m$^3$, the average number is about 35 Bq/m$^3$. Because the radon inside the detector had not been taken into account in the visual spectrum, so its radioactive should be simulation separately. Taking into account the radon and its sub-particles, considering the internal space of detector, the simulation result show that the background caused by radon is about 0.27cps/100/cm$^3$ Ge.

## 2.3  Comparison between the simulation and experimental data

A well-type HPGe detector from Canberra was used to verify this simulation method, it's relative efficiency is 40% and crystal volume is 170 cm$^3$, the probe is shielded with 15 cm thickness of oxygen-free copper, which was divided into six layers and each layer is 2.5 cm thick. The inner size of the copper shield is 35X35X55 cm$^3$. By adjusting the shield layers of the copper plates, the shielding effect can be tested, compare with these data, the simulation results cab be verified.

Firstly we measured the background spectrum in the laboratory without any shielding, analyzed it and then calculated the original spectrum of the visual source sphere. The calculated spectrum varies with the shapes of the probe and the radiuses of the visual source sphere. Fig. 2 is the experimental spectrum compared with the simulation spectrum. There are so many peaks, and most of them are the radiation of $^{214}$Bi, $^{208}$Tl and $^{228}$Ac, the daughters of $^{238}$U and $^{232}$Th. It is easy to find two main peaks, $^{40}$K at 1460.8 keV and $^{208}$Tl at 2614.7 keV. The Compton edges and Compton plateaus of these two peaks are obviously, while the escape peaks are small. Ignored some small peaks, the simulated and experimental data are well consistent. In addition, some simulated peaks offset a little comparing to experiment, which due to the non-linearity of the ADC conversion. In un-shielding HPGe simulation, the direct impact of cosmic rays with the HPGe probe is ignored , because counting rate of the integral background below 3 MeV induced by cosmic rays is only one five-hundredth of the $\gamma$ rays background.

Secondly, changed the shield copper thickness and simulated the shielding effect, note that the cosmic ray is considered for this simulation. At the same time, we tested the shielding effect of the same copper thickness corresponding to the simulation. Limited by the experimental conditions, the bottom side as the support device was set as 15 cm, only other five sides were tuned. Table 1 presents the counting-rate of integral background from 50 keV to 2.8 MeV. The background is obviously depressed when the thickness of shielding copper increases, while the muon background changes little. So the influence of cosmic ray dominates when the copper is thicker.

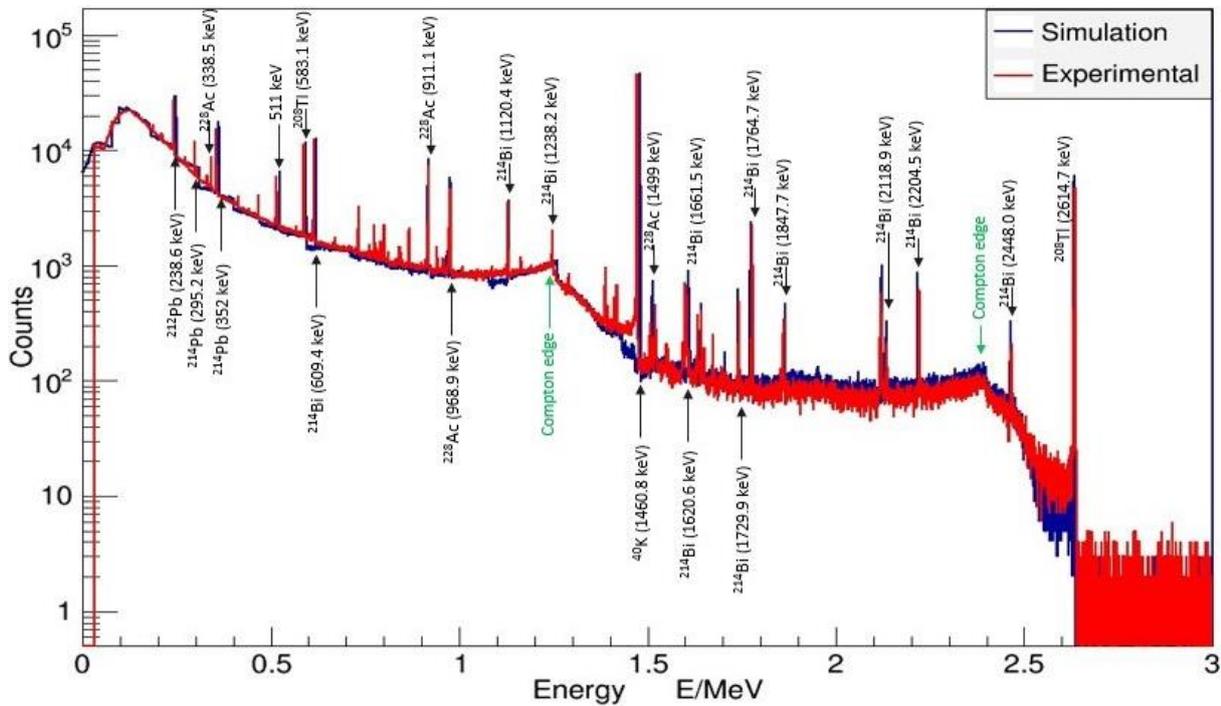

Fig.2 Experimental and simulated $\gamma$ rays spectrum of detected background from unshielded HPGe detector

The simulation results are lower than the data, and the difference is about 15% for 5.0 cm thick layer, 10% for other layers. The inconsistence come from the following aspects: the visual source spectrum is not entirely accurate, the unconsidered cosmic rays components could make the simulation results lower than test data, the simulation program codes would take in a little error when change the simulation conditions such as simulation times and initial particles. Even though there is a little difference, it can also be used to estimate the thickness of outer shield materials and find the proper dimensions of HPGe spectrometer.

Tab.1 The counting-rate of integral background (50keV~2.8MeV) comparison between data and simulation

| Copper Thickness (cm) | Simulation (cps/100 cm³Ge) | | | | Data (cps/100 cm³Ge) | Sim./Exp. Ratio |
|---|---|---|---|---|---|---|
| | Gamma | Muons | Radon | All | | |
| 2.5 | 32.87 | 0.351 | | 33.49 | 37.39 | 0.90 |
| 5.0 | 11.23 | 0.398 | | 11.9 | 13.95 | 0.85 |
| 7.5 | 4.049 | 0.386 | 0.27 | 4.705 | 5.13 | 0.92 |
| 10 | 1.494 | 0.374 | | 2.138 | 2.39 | 0.89 |
| 12.5 | 0.590 | 0.360 | | 1.220 | 1.38 | 0.88 |
| 15 | 0.326 | 0.34 | | 0.936 | 1.04 | 0.90 |

## 3. Simulation result of new detector

For surface HPGe detector, the background is just depressed down to about 1 cps/100cm³ Ge (50keV~2.8MeV) by 15-cm thick copper. Such level background is suitable for ordinary objects tests, but not for many low background samples. Therefore a new ultra low-background HPGe spectrometer as shown in Fig.3 is under construct. Compared with old spectrometer, the new one has three advantages: the active cosmic ray shielding detectors made of plastic scintillator, the lead layers encircled the copper, and the anti-Compton detector made of BGO crystal. With Geant4 simulation code, the dimensions of BGO crystal and the lead layers were optimized.

In the middle of the spectrometer is a four inch well-type HPGe detector from Canberra with a relative efficiency of 120%. The resolution is 2.1 Kev at 1.33 MeV. The nominal size of the Ge crystal is of 86.3 mm diameter and 85.8 mm height with a 4.89 mm front gap, and the aluminum shell is 1.5 mm thickness. Outer the probe, it is the anti-Compton detector made by 22 BGO crystals. Each BGO couples with a 2 inch low-background PMT for the single readout, and the inner volume of BGO crystals is $14 \times 14 \times 30$ cm$^3$. Surrounding the BGO crystals it is the walls made of Oxygen-free copper (15 cm thickness) and lead(10 cm thickness) to shield the γ background. The cosmic veto detector is made of plastic scintillator which installed outermost and covers five sides except the bottom.

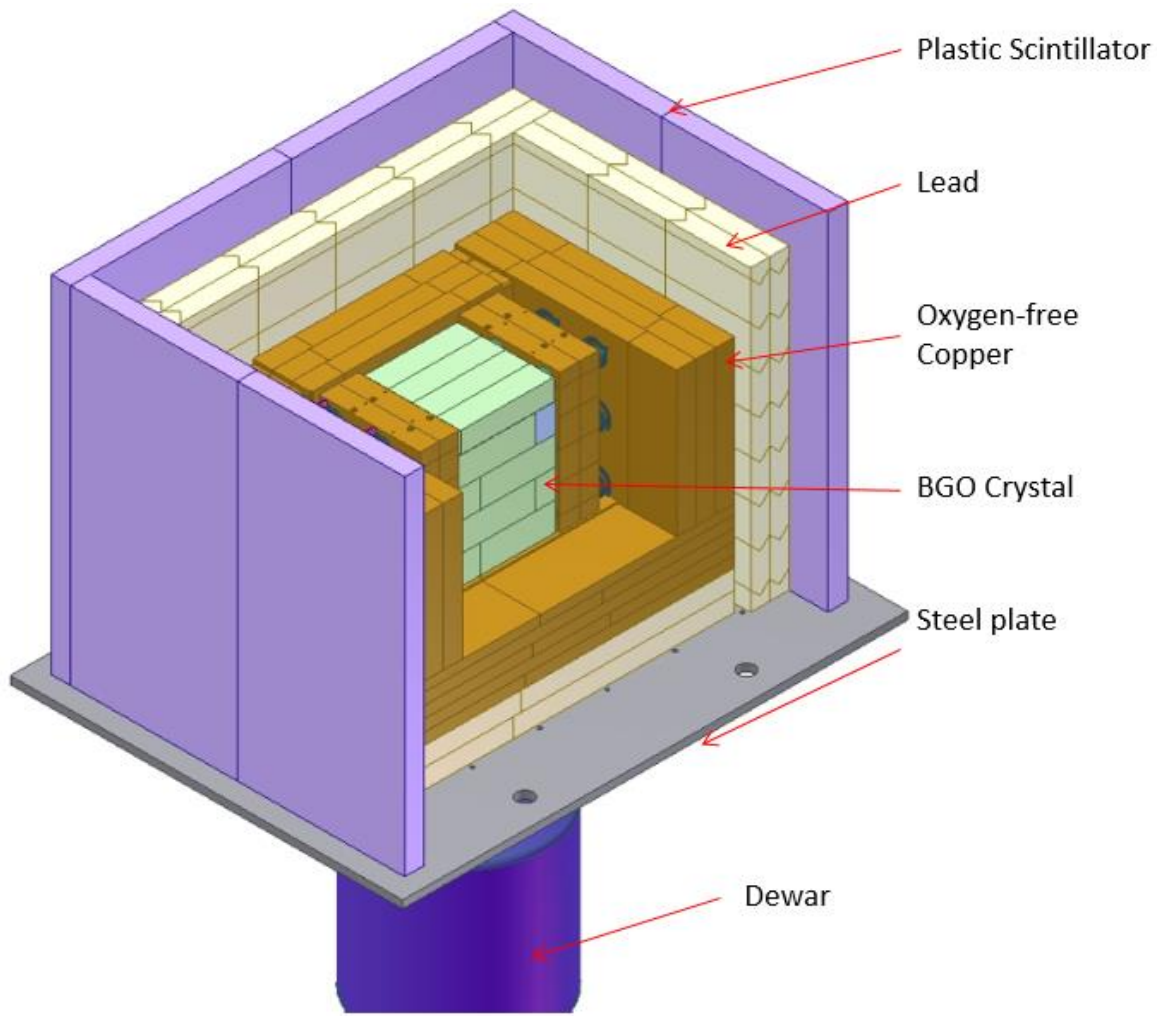

Fig.3 Sketch map of low background anti-Compton γ spectrometer

### 3.1 Anti-Compton detector simulation

Usually, the sample will be placed on the top surface of the probe for measurement. The radioactivity gamma rays from the sample come into the probe, and loss energy by the Compton Effect. If the γ particle is completely captured by the probe, a full energy peak will be obtained. While in most cases, because of the Compton scattering, γ particles penetrate through the probe depositing only a part of energy, which are Compton plateau events, they are useless for the radioactivity analysis, and sometimes even increases the errors. An anti-Compton detector surround the HPGe probe helps to veto those events and improve the detecting ability of the spectrometer.

Early Compton suppressors are usually based on NaI scintillator for high light output and large size. But NaI crystal is a little deliquescence and its radiation length is relative short. For low-background HPGe spectrometer, the bigger inner detector will increase the dimensions of shielding materials, and then result in more weight and background of the spectrometer. We choose $Bi_4Ge_4O_{12}$ (BGO) scintillator as the Compton depress detector, for its high density (7.13 g/cm$^3$). It offers a gamma-ray absorption coefficient about 2.5 times greater than that of NaI. Although the resolution of BGO is poorer, as a veto detector, the detection efficiency is more important, in this point, BGO is better than NaI and CsI crystals[9].

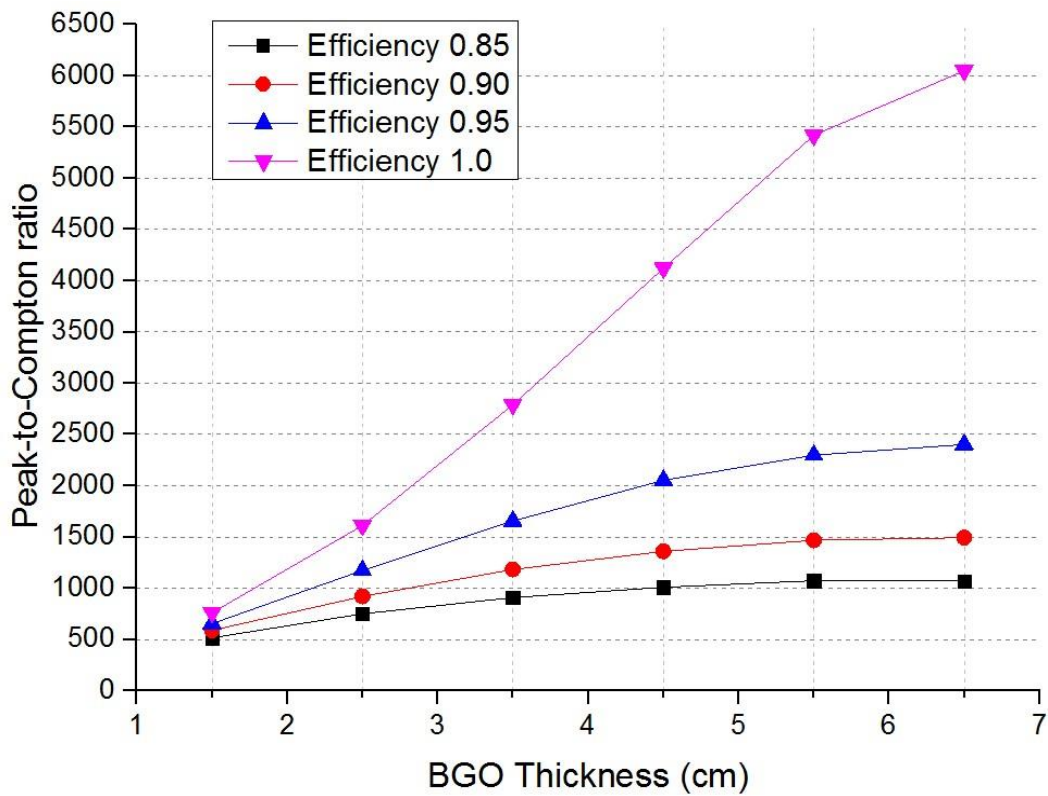

Fig.4 Peak-to-Compton ratio of $^{40}$K as a function of BGO thickness at different anti-coincidence efficiency

The anti-Compton detector was designed like a cubic shell by using 22 BGO crystals. The precise dimensions and relative position of BGO crystals to HPGe probe were described in simulation. A point $^{40}$K source was placed 2.5 cm above the top surface of HPGe, and the energy resolution is set about 2.1 keV @ 1460.8 keV. One way to improve peak-to-Compton ratio is to increase the detecting efficiency of BGO by using thicker and bright crystals, another is to advance the anti-coincidence efficiency. Fig.4 shows the simulation results considering the two factors. It is obvious that the peak-to-Compton ratio increases with the increase of BGO thickness, and 5.5 cm is a key point, at which the Compton depression improved little after. For the ideal anti-coincidence efficiency 1.0 (100%), the upward trend also slows down after 5.5 cm, and the peak-to-Compton ratio is almost the same at 5.5 and 6.5 cm thickness when the anti-coincidence efficiency is below 0.9 (90%). The electronics system affects the anti-coincidence efficiency a lot in the real measurement. By adjusting the gate width and relative time corresponding to the probe signals, the proper working status can be obtained.

### 3.2 Lead shield simulation

Because the low-background lead is so rare, we choose the ordinary lead plus oxygen-free copper to shield the γ rays. Oxygen-free copper has lower radioactivity than other materials, placed in the inner of the lead can shield the background from lead material. The most important performance index of the new low-background HPGe spectrometer is that the integral background counting-rate from 50keV to 2MeV is lower than 0.01 cps/100 cm$^3$ Ge crystal, so the environment γ-ray, as one source of background, must be shielded to tenth or fifth of the index. Table 2 shows the simulation results of environment gammas and cosmic muons background after shielding with different thick lead, 10cm lead reduced the gamma background to 0.0024 cps. As to the Muons background, the simulation result at 10-cm lead is 0.2885. According to the experience, five sides covered by 5cm thick plastic scintillator can reduce more than 97% of the cosmic ray induced background, table 2 show that 10cm thick lead with 5cm cosmic ray veto detector is better.

Tab.2 The simulation counting-rate of integral background(50keV~2.8MeV)shielded by lead.

| Lead Thick(cm) | Simulation(cps/100cm$^3$Ge) | | | |
|---|---|---|---|---|
| | Gamma | Muons | Cosmic Veto(97%) | All |
| 2 | 0.1139 | 0.3262 | 0.00979 | 0.1237 |
| 4 | 0.0351 | 0.3129 | 0.00939 | 0.0445 |
| 6 | 0.0119 | 0.3087 | 0.00926 | 0.212 |
| 8 | 0.0032 | 0.397 | 0.00891 | 0.121 |
| 10 | 0.0024 | 0.2885 | 0.00865 | 0.0111 |
| 12 | 0.0008 | 0.2757 | 0.00827 | 0.0091 |

### 4. Summary

Low-background HPGe detectors usually use a lot of shielding materials, which leads to a heavy and difficult disassembly device. Geant4 simulation is a good way to evaluate the design in advance, and optimize the dimensions of the detector. The background simulation method presented in this study is suitable for the shielding materials simulation, because the visual source spectrum generated through the actual measurement background is more accurate. For our new HPGe spectrometer, the simulation results show that 5.5-cm thick BGO crystal is the optimal choice for the anti-Compton detector, and 15-cm oxygen-free copper plus 10-cm lead could depress the gamma integral background to 0.0024 cps/100 cm$^3$ Ge crystal, which is one fifth of the design index. In addition, as to the real experiment, the devices placed in the copper such as cables and PMTs are the fatal background sources to the HPGe probe, the low-radioactivity types of these devices should be chosen for HPGe detectors, also continuous pure nitrogen filling in the detector could depress the radon radioactivity mostly.

### References

1. R J McDonald, A R Smith, D L Hurleyet et al,Journal of Radio analytical and Nuclear Chemistry,1998,234: 33-36
2. Ph Hubert, Nucl.Instrum.Methods A,2007, 580: 751-755
3. S Agostinelli, J Allison, K Amako et al, Nucl.Instrum.Methods A, 2003, 506: 250-303
4. Wang Dian-Rong et al, Physica Energiae Fortis ET Physica Nuclearis, 1983, 7: 136-141
5. Second Experimental Group of the Institute of High Energy Physics, Physica Energiae Fortis ET Physica

Nuclearis,1983, 7: 401-407
6. R.I. Obed et al, Journal of Environmental Radioactivity, 2011,102: 1012-1017
7. R.Antanasijevi et al, Radiation Measurements, 1999, 31:371-374
8. V.Udoviic et al, Radiation Measurements, 2009, 44:1009-1012
9. Stromswold D C, IEEE.Trans.Nucl.Sci, Ns-28, 290(1981)